\documentclass[runningheads]{llncs}
\usepackage{graphicx, amssymb, amsmath}
\usepackage{multirow, color}
\usepackage[font=small,skip=5pt]{caption}
\newcommand{\duy}[1]{\textcolor{black}{#1}}
\newcommand{\hg}[1]{\textcolor{black}{#1}}

\begin{document}
%



\title{Improving the Accuracy of Transaction-Based Ponzi Detection on Ethereum}

\titlerunning{Ponzi Schemes Detection on Ethereum}
%
%

\author{
Phuong Duy Huynh\inst{1}\and
Son Hoang Dau\inst{1} \and
Xiaodong Li\inst{1} \and
Phuc Luong\inst{2} \and
Emanuele Viterbo\inst{2}
}
\authorrunning{P. D. Huynh et al.}
\institute{RMIT University, Melbourne VIC 3000, AU \and Monash University, Clayton VIC 3800, AU
}

\maketitle              
\begin{abstract}
The Ponzi scheme, an old-fashioned fraud, is now popular on the Ethereum blockchain, causing considerable financial losses to many crypto investors. A few Ponzi detection methods have been proposed in the literature, most of which detect a Ponzi scheme based on its smart contract source code. This \textit{contract-code-based} approach, while achieving very high accuracy, is \textit{not robust} because a Ponzi developer can fool a detection model by obfuscating the opcode or inventing a new profit distribution logic that cannot be detected. On the contrary, a \textit{transaction-based} approach could improve the robustness of detection because transactions, unlike smart contracts, are harder to be manipulated. However, the current transaction-based detection models achieve fairly \textit{low accuracy}. 
In this paper, we aim to improve the accuracy of the transaction-based models by employing \textit{time-series features}, which turn out to be crucial in capturing the life-time behaviour a Ponzi application but were completely overlooked in previous works. We propose a new set of 85 features (22 known account-based and 63 new time-series features), which allows off-the-shelf machine learning algorithms to achieve up to $30\%$ higher F1-scores compared to existing works.
\end{abstract}

\section{Introduction}
Since the birth of Bitcoin in 2008~\cite{nakamoto2008bitcoin},
the blockchain technology has grown exponentially and revolutionised the way currencies and digital assets are transferred. 
Thanks to its inherent decentralisation, anonymity, and immutability, a blockchain provides better tampering resistance, robustness, privacy protection, and cheaper turn-around costs compared to traditional financial systems~\cite{8246573,ZHAO2020102355}. Apart from digital finance applications, Turing-complete smart contracts introduced first by Ethereum~\cite{Ethereum2013} and then by other similar blockchain platforms allow developers to implement sophisticated logic on the chain, further expanding the applicability of the technology to many other sectors including supply chains \cite{dutta2020blockchain,korpela2017digital,casado2018blockchain}, data sharing \cite{theodouli2018design,jaiman2020consent},
and the internet of things \cite{panarello2018blockchain,dorri2017towards,novo2018blockchain}.

In recent years, crypto-crowdfunding via initial coin offerings (ICOs) has become a major fundraising method used by many businesses~\cite{MORKUNAS2019295}, providing an attractive alternative to traditional stock exchanges.
By the end of December 2023, the global market capitalization of blockchains had reached a staggering amount of over \$1.7 trillion with more than 2.2 million different cryptocurrencies \cite{CoinmarketCap2023}. 
However, this phenomenal success of the blockchain technology in digital finance has also led to a rising number of cybercrimes. 
Smart-contract-supporting blockchains have now become a paradise for a plethora of devastating financial scams, such as Ponzi, Honeypots, Trapdoor, Phishing,
and Rug Pull~\cite{EtherScam2019,huynh2023programming}.


A Ponzi scheme is a scam that promises high returns to investors by using the funds from newcomers to pay earlier investors. This scam collapses when few or no new investors join, making most investors, except for the early ones and the scheme owner, lose their investment. According to Chainanalysis's 2021 Crypto Crime Report~\cite{EtherScam2021}, from 2017 to 2020, most blockchain frauds were Ponzi schemes, which accounted for nearly \$7 billion worth of cryptocurrency in 2019, more than double of all other scams in 2020. The development of Ponzi schemes on Ethereum\footnote{In the scope of this work we focus on Ponzi schemes on Ethereum only. For Bitcoin-based Ponzi schemes, please refer to~\cite{VasekMoor2015,BartolettiPesSerusi2018}.} have attracted some attention from the research community. 
The first work was done by Bartoletti \emph{et al.} \cite{BARTOLETTI2020259}, who analysed Solidity \textit{source codes} of smart contracts and proposed four criteria to identify a Ponzi scheme (their paper first appeared on Arxiv in 2017). They classified Ponzi schemes on the Ethereum chain into four different types according to their money distribution logic. 
They also constructed the very first Ponzi dataset on Ethereum, consisting of 184 Ponzi contracts by manually inspecting their source codes. To automatically detect Ponzi schemes, a number of detection models using machine learning methods~\cite{chen2018detecting,chen2019exploiting,jung2019data,fan2020expose,wang2021ponzi,zhang2021detecting,ibba2021evaluating} and symbolic execution techniques~\cite{chen2021sadponzi} have been developed in the literature. 

Most of the machine learning approaches employed both transaction-based features (e.g., account features) and contract-code-based features (e.g., \textit{opcode} features) in their models to improve the detection accuracy. 
However, a \textit{contract-code-based approach}, while capable of achieving very high accuracy, is \emph{not robust}. \textit{First}, the Solidity source codes of a majority of contracts on Ethereum are not available~\cite{zhou2018erays}. \textit{Second}, a
Ponzi operator can fool a contract-code-based detection model by obfuscating the opcode (see~\cite[Section 7.2.1]{chen2021sadponzi}) or inventing a new profit distribution logic that cannot be detected (see~\cite[Section 8]{chen2021sadponzi}). 
A \textit{transaction-based approach} could improve the
robustness of detection because transactions, unlike smart contracts, are harder to be manipulated (see Section~\ref{subsec:related_work}). However, the current transaction-based detection models~\cite{chen2018detecting,chen2019exploiting,jung2019data} achieve fairly poor performance
, e.g. 44\% and 68.8\% (see the discussion in Experiment 1, Section~\ref{subsec:experiments}).

In this work, we aim to develop more \textit{robust} and \textit{accurate} detection models that only rely on transaction data. To this end, we first retrieved from the XBlock-ETH repository~\cite{xblockweb} all transactions associated with 1395 labeled applications provided in~\cite{chen2021sadponzi}. We then analysed the data to capture the way Ponzi applications work. We observed that Ponzi and non-Ponzi applications have distinctive behaviours and characteristics and more importantly, that the \textit{time} factor, which has been overlooked in most studies, is crucial in identifying a Ponzi application.
For example, the balance of a Ponzi account may have several ``cliffs" in its lifetime, which occurred when funds were gradually accumulated and then paid out to one or several investors. As another example, a Ponzi application often has shorter lifespan with a peak transaction volume around its creation time, and then few to none transaction after that.   
Based on such intuition, we designed a new feature list that consists of existing features and novel \textit{time-series features} that capture the behaviours of an application throughout its lifetime.

To evaluate the effectiveness of the proposed list of features, we ran different machine learning algorithms on this list and on the existing lists of features used by other transaction-based models~\cite{chen2018detecting,jung2019data}, treated as the baselines. 
Analysing the list of important features from the best performing model (LGBM), we found that \textit{time-series features} indeed contributed 
to the improvement of the F1-score of the model. The improvement is up to \duy{5.7\%} as compared to models that only used \textit{account features}. Furthermore, using LGBM's feature importance, we trimmed our original feature set of size 545 to obtain a much smaller one consisting of the 85 most important features (for LGBM). It is remarkable that the trimmed feature set outperforms the original one for \textit{all} performance metrics (Accuracy, Precision, Recall, F1-score, and running time) in \textit{all} five off-the-shell machine learning algorithms.
Especially, we observed a sharp increase in F1-score when using the new list of 85 features (22 known features and 63 time-series features) compared to when using the existing lists from~\cite{chen2018detecting} and~\cite{jung2019data}. More specifically, our model achieved 11\% higher F1-score compared to~\cite{jung2019data} (Random Forest) and 30\% higher F1-score compared to~\cite{chen2018detecting} (XGBoost). Last but not least, we demonstrated that our approach can also detect, with high accuracy, \textit{new} types of Ponzi schemes that were not present in the training dataset.

The rest of the paper is organised as follows. In Section~\ref{subsec:background}, we introduce the background knowledge and discuss related work. We describe how to construct time-series features in Section~\ref{subsec:framework} and demonstrate the effectiveness of our new feature set via extensive experiments in Section~\ref{subsec:experiment}.
We conclude the paper in Section~\ref{subsec:conclusion}.

\section{Background}
\label{subsec:background}

\vspace{-5pt}
\subsection{Ethereum in a Nutshell}
\label{subsec:ethereum}
Ethereum is the second most popular blockchain after Bitcoin in terms of market capitalization \cite{CoinmarketCap2022}. It is also the largest platform that provides a decentralised virtual environment (i.e., Ethereum Virtual Machine or EVM for short) to execute smart contracts \cite{NSzaboSmartContract1996}. In 2022, the Ethereum chain reached 15 million blocks with over 1.5 billion transactions \cite{Etherscan}. \textit{Smart contracts} on Ethereum are executable programs that run automatically when their trigger conditions are met. Those contracts can be implemented using an object-oriented and high-level language called Solidity~\cite{Solidity}. Contract \textit{source codes} are then compiled into \textit{bytecodes}, which can be represented as low-level human-readable instructions - \textit{opcodes} \cite{GWoodEthereum2014}. After that, the bytecodes are launched onto EVM. Once a contract is deployed, it cannot be modified by anyone. Moreover, any activity in the life cycle of a contract
must be triggered by a transaction. Therefore, any interaction `from' or `to' a contract is recorded as a transaction and stored on the blockchain permanently. In other words, in Ethereum, a transaction is a key unit that involves all activities of a contract.

\subsection{Ponzi Schemes on Ethereum}
\label{subsec:ponzi}
The blockchain technology, although has the potential to revolutionise the way traditional businesses work~\cite{8970310}, 
also creates a golden opportunity for cybercriminals, resulting in the migrations of many financial scams to the blockchain platforms~\cite{EtherScam2019}. Among blockchain scams, Ponzi schemes~\cite{ARTZROUNI2009190} were the most popular from 2017 to 2020. 
In hindsight, this is not a surprise because Blockchain's inherent properties, i.e., automation, transparency, immutability, and anonymity create an ideal environment for this scam to grow~\cite{LI2020841}. 

In layman term, Ponzi schemes are scams often camouflaged as high-return investment programs that use the funds from new investors to pay existing ones. 
With no real project behind and no intrinsic value, a Ponzi scheme will collapse when there are not enough new investors joining and/or the payment commitment can no longer be fulfilled. A more official and authoritative definition of Ponzi schemes is given by the U.S. Securities and Exchange Commission~\cite{PonziDefUS}. 
At the heart of each Ponzi scheme is a money redistribution mechanism.
\hg{Bartoletti \emph{et al.}~\cite{BARTOLETTI2020259} classified Ponzi schemes on Ethereum into four different categories based on their redistribution mechanisms, including Chain-shaped, Tree-shaped, Handover, and Waterfall schemes (see Appendix~\ref{app:Ponzi_types})}.

\vspace{-5pt}
\subsection{Related Works}
\label{subsec:related_work}
Existing Ponzi detection models can be divided into two groups, depending on whether they rely on smart contract codes or on the transactions.

\textbf{Contract-Code-Based Approaches:} 
Bartoletti \emph{et al.} \cite{BARTOLETTI2020259} first proposed four criteria to detect a Ponzi application by inspecting their contract source codes. 
However, it turns out that the Solidity source codes of 77.3\% contracts on Ethereum are not available~\cite{zhou2018erays}. 
To tackle this drawback and to detect Ponzi automatically, many researchers built Ponzi detection tools based on the frequency distribution of operation codes (opcodes), which are always available on the Ethereum chain. Chen \emph{et al.}~\cite{chen2018detecting,chen2019exploiting} proposed an automatic detection tool on opcode features using machine learning models. Their experimental results showed that the detection models using opcode features achieved greater performance than those using account features, which were aggregated from transactions. 
Fan \emph{et al.} \cite{fan2021spsd} pointed out that an imbalanced dataset caused an over-fit in previous works~\cite{chen2018detecting,chen2019exploiting}. \hg{To improve data quality and the detection accuracy, they proposed a data enhancement method 
that expanded the dataset and eliminated the imbalance
using data sampling techniques. 
Wang \emph{et al.}~\cite{wang2021ponzi} adopted a deep learning technique to build a more accurate detection tool and also used oversampling techniques (Smote and Tomek) to deal with an imbalanced dataset.
Jung \emph{et al.}~\cite{jung2019data} 
and Sun \emph{et al.}~\cite{sun2020early} focused more on crafting better representative features than improving 
data quality.}

\hg{A common drawback of all previously mentioned studies is the \textit{lack of robustness}.} 
As pointed out by Chen~\emph{et al.}~\cite{chen2021sadponzi}, scammers can use code obfuscation techniques~\cite{BiAn} to counter those detection models that rely on opcode features (see~\cite[Section 7.2.1]{chen2021sadponzi}). 
\hg{For example, a contract code can be manipulated or modified to change the opcode occurrence frequency.
Chen \emph{et al.}~\cite{chen2021sadponzi} also proposed in their work a new detection tool called SADPonzi, which was built upon a semantic-aware approach and achieved 
100\% \texttt{Precision} and \texttt{Recall}.} 
SADPonzi was proven to be more robust than the current opcode-based method when facing code obfuscation techniques. 
More specifically, it can detect a Ponzi contract by comparing the extracted semantic information of its bytecode and the predefined semantics of four known Ponzi schemes. 
However, the approach of SADPonzi requires a domain expert to analyse a Ponzi application's operational logic to build the corresponding semantic pattern, which can be costly to put into practice. On top of that, as also mentioned by the authors, SADPonzi can only effectively detect known Ponzi types with predefined semantics, and may fail to detect a new Ponzi variant (see~\cite[Section 8]{chen2021sadponzi}). 

\textbf{Transaction-Based Approaches:}
Transactions are records that save historical activities between an application and its participants. 
Transaction data was used in some existing works~\cite{BARTOLETTI2020259,chen2018detecting,chen2019exploiting} to capture the differences between Ponzi and non-Ponzi applications.
Detection tools based on transaction data are more resilient to scammers' countermeasures because transaction information cannot be modified or deleted from the chain.
Although scammers can add transaction records, they cannot manipulate transaction data as freely as they can with smart contract's source code and opcodes for two reasons. \textit{First}, any participant, not just the creator, can create transactions, which is not under the control of the contract creator. \textit{Second}, the cost to create a transaction on the chain is expensive (approximately \$14.26 on average per transaction~\cite{EtherTxsFee}). 
\hg{These factors prevent the Ponzi scammer from manipulating their transaction data arbitrarily to elude detection, e.g. by flooding the system with fake transactions.}

\hg{Despite several advantages, existing transaction-based models~\cite{chen2018detecting,chen2019exploiting,chen2021sadponzi} suffered from \textit{low classification accuracy}, achieving F1-scores around 44\%-69\% only. The key reason for their mediocre performance, based on our analysis, could be due to the fact that existing works have completely missed the \textit{time dimension} when building their models.
We note that some studies \cite{chen2018detecting,chen2019exploiting,jung2019data,wang2021ponzi}, in order to improve the detection accuracy, integrated account features with opcode features. However, such a hybrid approach also inherits the shortcomings of the contract-code-based approach. 
\textit{In this work}, we explored the temporal behaviour of Ponzi applications and introduced \textit{time-series features} alongside existing account features, aiming for \textit{both} robustness and accuracy in detection.}

\vspace{-5pt}
\section{Transaction-Based Features Extraction}
\label{subsec:framework}
\vspace{-5pt}

\subsection{Data Collection}
\label{subsec:data_collection}

\hg{We refined the dataset of labeled Ponzi and non-Ponzi addresses provided in the SADPonzi paper~\cite{chen2021sadponzi} by first downloading and extracting transaction histories of these contracts from the XBlock-ETH repository~\cite{Zheng_2020,xblockweb}.} We then filtered out unsuccessful transactions which failed for various reasons such as insufficient gas (a required fee to successfully conduct a transaction) or errors in the contract codes. 
\hg{We also discarded applications with no transactions or having lifetime (the period from the recreation time to the last transaction) shorter than one day. 
These are outliers, whose behaviours do not resemble the whole group.} Even if such an application was a Ponzi, it was also a failed scam.
Therefore, removing those applications is important to build a clean dataset, especially for a transaction-based approach. 
Our refined dataset contains \textbf{1182} non-Ponzi and \textbf{79} Ponzi applications. The Ponzi types statistics are displayed in Table~\ref{tab:ponzi_types}.

\vspace{-20pt}
\begin{table}[htb!]
\centering
\caption{\hg{Ponzi types (see Appendix~\ref{app:Ponzi_types})} statistics for our refined dataset.} 
\label{tab:ponzi_types}
\begin{tabular}{|l|c|c|}
\hline
\textbf{Ponzi type}     &   \textbf{Number of applications}       &     \textbf{Percentage}    \\ \hline
Chain-shaped     &         68                     &           86\%        \\ \hline
Tree-shaped      &         1                      &           1.3\%        \\ \hline
Handover         &         1                      &           1.3\%        \\ \hline
Waterfall        &         4                      &           5\%        \\ \hline
Other            &         5                      &           6.4 \%        \\ \hline
\end{tabular}
\vspace{-25pt}
\end{table}

\subsection{The Importance of Temporal Behaviour in Ponzi Detection}
\label{sec:data_analysis}

\hg{In this section, we investigate how Ponzi and non-Ponzi applications work differently with respect to their temporal behaviours.
To this end, we chose to} analyse the historical transaction data of DynamicPyramid, a representative Ponzi contract\footnote{0xa9e4e3b1da2462752aea980698c335e70e9ab26c (DynamicPyramid's address)}. This is a chain-shaped scheme, the most popular type, which constitutes 86\% of all known Ponzi contracts. 
In general, different types of applications have different transaction behaviours, and understandably, Ponzi applications have unique behaviours that are different from non-Ponzi ones. 
In our analysis, we compare the representative applications of the two groups to demonstrate their potential differences regarding temporal behaviours. \hg{A comprehensive comparison between different types of Ponzi and all different types of non-Ponzi, while valuable, would be an overkill for our purpose, and hence, out of the scope.} 

\begin{figure}[htb!]
\centering
\includegraphics[width=12cm]{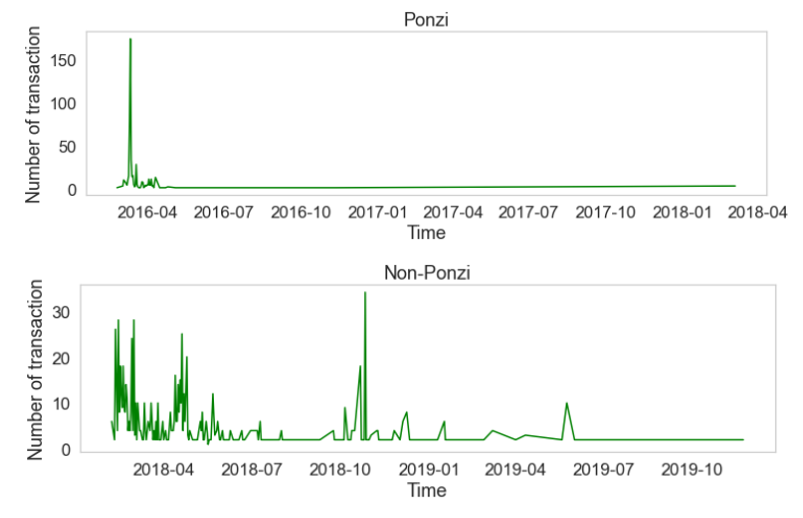}
\caption{Daily transaction volumes of a Ponzi (DynamicPyramid) and a non-Ponzi applications. The Ponzi application had a shorter lifespan with a peak transaction volume concentrating in the first month followed by almost no activities. By contrast, the non-Ponzi application had more regular activities throughout its long lifespan.}
\label{fig:daily_tx_volume}
\vspace{-18pt}
\end{figure}

\textbf{Transaction volumes.} We start our analysis by comparing the \textit{transaction volumes} of a Ponzi application (DynamicPyramid) and a non-Ponzi application\footnote{0xb2c3531f77ee0a7ec7094a0bc87ef4a269e0bcfc (a non-Ponzi contract address)}. 
The transaction volume of an application measures the daily number of associated transactions. As observed in Fig.~\ref{fig:daily_tx_volume}, DynamicPyramid 
had a shorter lifespan with a peak transaction volume concentrating in the first month followed by almost no activities.
In comparison, the non-Ponzi application had more regular activities throughout its long lifespan. \hg{The reason is that a Ponzi application, although often introduces itself as a potential project with a high investment return promise, has 
no actual project behind.} Thus, participants of Ponzi applications often \hg{participated} 
actively at the beginning as \hg{early investors got paid regularly. 
However, as time goes by, fewer investors got paid and more will start leaving, leading to the unavoidable collapse.}

\begin{figure}[htb!]
\centering
\includegraphics[width=12cm]{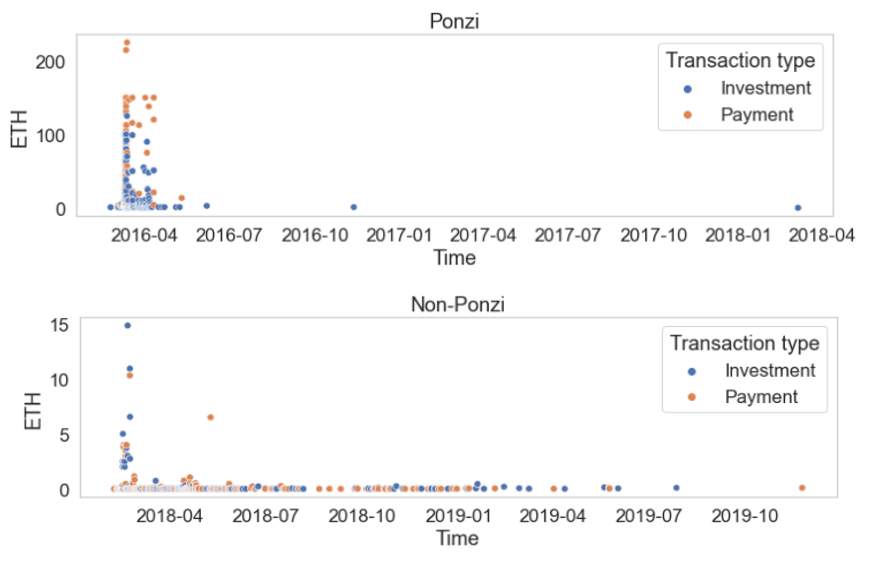}
\caption{Investment and payment activities of a Ponzi (DynamicPyramid) and a non-Ponzi applications. Several lower investments (blue dots) were followed by a higher payment (orange dot) in the Ponzi application, which demonstrates the funds accumulation before a payment to an investor can be made.}
\label{fig:project_activities}
\end{figure}

\textbf{Investment versus payment activities.} Pushing our analysis one step further, we break down transactions into two different 
types, namely, \textit{investments} and \textit{payments}. An investment refers to a transaction sending ETH from an investor to an application, whereas a payment refers to a transaction from an application that pays ETH to an investor. As demonstrated in Fig.~\ref{fig:project_activities}, payments (orange dots) and investments (blue dots) of the Ponzi application concentrated only in the first month. Moreover, each orange dot was preceded by some blue dots of smaller ETH amounts. This is because the examined Ponzi application, a chain-shaped scheme, must gather sufficient new investment funds before making a payment to a single participant. After this intensely active period, the number of payments decreased and finally disappeared. 
This happened because the application's balance was no longer enough to make any new payment \hg{despite having a few new investments coming in}. 
\hg{On the contrary, both investment and payment activities spread out over the lifespan of the non-Ponzi application.}

\begin{figure}[htb!]
\centering
\includegraphics[width=11cm]{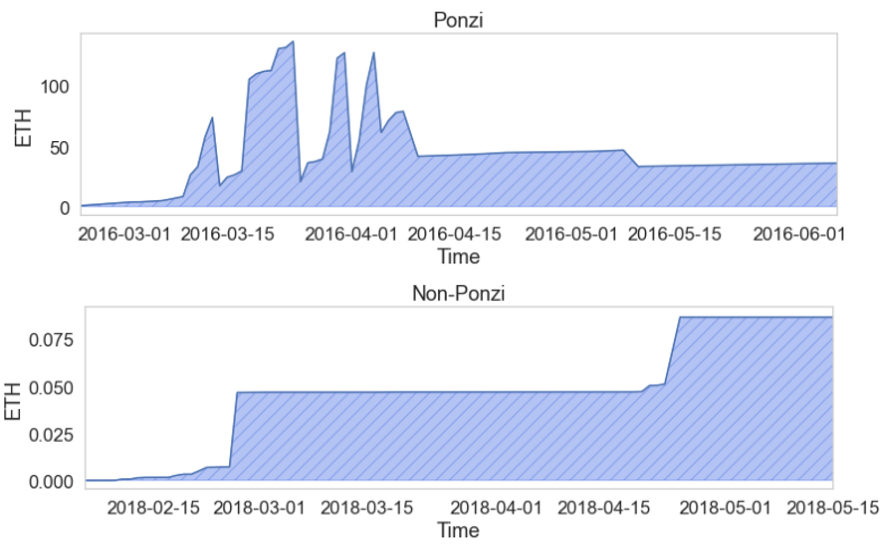}
\caption{\hg{Application balances (in the first four months after launch) of a Ponzi application (DynamicPyramid) and a non-Ponzi application. As observed, the chart of the Ponzi contract had a number of ``cliffs'' while that of the non-Ponzi contract had none.}} 
\label{fig:contract_balance}
\end{figure}

\textbf{Application balance.} The \textit{balance} of an application is the amount of ETH in the application at a time. How the balance varies as time goes by can indicate the type of application. As demonstrated in Fig.~\ref{fig:contract_balance}, the balance of the Ponzi application (Dynamic Pyramid) often rose gradually (investments), and after a while, dropped dramatically, generating a ``cliff'' (payment). The reason is that the balance was gradually accumulated from the investments until reaching the amount that the application had committed to pay to a particular investor when they joined the application. Once the desired balance was reached, the promised profit was immediately paid to the corresponding investor.

\vspace{-15pt}

\subsection{Transaction-Based Features Extraction}
\label{subsec:features_agg}

\hg{We classify transaction-based features into two types: \textit{account features} and \textit{time-series features}. 
Transaction-based detection models proposed so far in the literature only used account features~\cite{chen2018detecting}, which capture general statistics of the transactions associated with the application}, e.g. the total/average investment amount, the final balance of the contract, or the maximum number of payments to an investor. \hg{As mentioned in Section~\ref{sec:data_analysis}, using only account features led to rather poor prediction performance (low F1-scores). 
To improve transaction-based Ponzi detection models, it is essential to employ both account features and time-series features. 
We discuss in detail below how to extract these features, especially the new time-series ones.}

\textbf{Account features:} \hg{This type of features captures general information about the contract of interest and} has been widely used in previous studies \cite{chen2018detecting,chen2019exploiting,jung2019data,wang2021ponzi}. 
More specifically, general statistic metrics such as average, count, sum, standard deviation, and Gini coefficient
\cite{gini} can be extracted from the set of all relevant transactions to aggregate account features. Although account features are insufficient to capture all behaviours of the Ponzi scheme, they are still very useful in revealing 
Ponzi's working logic. For example, the Gini coefficient of the number of payments can demonstrate an inequality in money distribution, or the final balance of the application indicates whether the investment funds have been distributed all to investors. \hg{Therefore, 
we still include in our list 29 account features introduced earlier in the literature~\cite{chen2019exploiting,chen2018detecting,jung2019data,wang2021ponzi} (listed in Appendix~\ref{Ap:afeatures}).}

\textbf{Time-series features:} 
As discussed earlier, time-series features play an important role in identifying Ponzi applications. Unlike account features, they capture the behaviours and activities throughout the application's lifetime. 
To aggregate time-series features, we \textit{first} partitioned our transactions into several \textit{time intervals} (\hg{with interval length of either 12, 24, or 48 hours}) 
and built 43 \textit{time series} (see Appendix~\ref{Ap:ifeatures} for the complete list), which measure various aspects of the transactions. 
These times series form three dimensions, namely, the contract address, the interval, and the data value for that contract in that interval (e.g. account balance).  
We then use a dimensionality-reduction technique to capture various  characteristics (e.g, mean, entropy, spikiness) of aggregated time-series and mapped the 3-D data into the 2-D space to produce the final time-series features. The creation steps of time-series are depicted clearly below:

    (1) For a fixed time duration $T$, we split our transaction data into $N$ time intervals of length $T$ each where $N \triangleq \lceil\texttt{life\_time} / T\rceil$. In our study, we have chosen 3 different values of T to study, which are 12, 24, and 48 hours.
     
    (2) Based on the \texttt{timestamp} field, we assigned each transaction to its corresponding interval.
    
    (3) We created \hg{a comprehensive list of 43 time-series 
    to represent all} activities that occur during the application's lifetime. 
    Thus, for each application, the time-series can be represented as a 2-D matrix of size $N\times 43$. Lastly, if the dataset has $M$ applications, the time-series data extracted from the whole dataset can be represented as a 3-D array with size $M\times N\times 43$.

Finally, to generate time-series features, we applied a dimensionality-reduction technique~\cite{hyndman2015large} to compress the time-series data. In particular, we employed a finite set of 12 statistical measures (see Appendix~\ref{Ap:tsmeasures}) proposed in \cite{deng2013time,fulcher2014highly,hyndman2015large,wang2006characteristic} to capture the global information of the 43 time-series and compressed the 3-D time-series data down to a 2-D $M\times 516$ matrix (note that $516 = 43\times 12$). The Python codes for time-series features aggregation are available at~\cite{git_anonymous}.
\vspace{-10pt}
\section{Experiments}
\label{subsec:experiment}

\subsection{Machine Learning Models}
To measure the effectiveness of our proposed 
set of features (account and time-series features), we reused classification methods employed in the previous studies~ \cite{jung2019data,chen2018detecting}, namely, \textbf{Random Forest (RF)}~\cite{svetnik2003random} and \textbf{XGBoost (XGB)}~\cite{chen2016xgboost}. In addition, other well-known classification methods such as \textbf{K-nearest neighbour (KNN)} \cite{cover1967nearest}, \textbf{Support vector machine (SVM)}  \cite{hearst1998support}, and \textbf{LightGBM (LGBM)} \cite{ke2017lightgbm} were also included in our experiments in order to find the most suitable classification model for the problem. RF uses the bootstrap resampling technique to generate different training decision trees from the original dataset and a prediction is made by aggregating the predictions from these trees. This algorithm works effectively in several domains \cite{rokach2016decision} including fraud detection~\cite{bhattacharyya2011data}. XGB is a gradient-boosting based algorithm that creates gradient-boosted decision trees in sequential form and then groups these trees to form a strong model. Unlike RF, the result of XGB is the prediction of the last model, which addressed data misclassified from previous models. KNN is a non-parametric classifier that uses proximity to estimate the likelihood of a data point belonging to one group. SVM is a classic algorithm that has been widely applied in binary classification or fraud detection problems, which 
performs classification by establishing a hyperplane that enlarges the boundary between two categories in a multi-dimensional feature space. LGBM is also a gradient-boosting-based algorithm similarly to XGB. However, LGBM grows a tree vertically (leaf-wise) while XGB grows trees horizontally (level-wise). With leaf-wise algorithms, LGBM is often more accurate and faster than other gradient-boosting-based algorithms. The Scikit-learn python library\footnote{https://scikit-learn.org/stable/} was used for \textbf{RF} (default parameters), \textbf{KNN} (default parameters), \textbf{SVM} (default parameters with ``poly" kernel), and \textbf{XGB} (default parameters without using label encoder). The Microsoft LightGBM Python library\footnote{https://github.com/microsoft/LightGBM} was used for \textbf{LGBM} (with default parameters).

\subsection{Evaluation Metrics, Model Structure and Experiment Setting}

\hg{To evaluate the performance of our models, we use standard prediction metrics including \textbf{Accuracy}, \textbf{Precision}, \textbf{Recall}, and \textbf{F1-score}, which are calculated based on the numbers of true positives (\texttt{TP}), true negatives (\texttt{TN}), false positives (\texttt{FP}), and false negatives (\texttt{FN}) as follows.
$\textbf{Accuracy}\triangleq (\texttt{TP}\hspace{-1pt} +\hspace{-1pt} \texttt{TN})/(\texttt{TP} \hspace{-2pt}+\hspace{-2pt} \texttt{FP} \hspace{-2pt}+\hspace{-2pt} \texttt{TN} \hspace{-2pt}+\hspace{-2pt} \texttt{FN})$ is the fraction of correct predictions, $\textbf{Precision}\triangleq \texttt{TP}/(\texttt{TP + FP})$ is the fraction of the actual Ponzi applications out of all the predicted Ponzi by the method, $\textbf{Recall} \triangleq \texttt{TP}/(\texttt{TP + FN})$ is the fraction of detected scams among all actual scams, and $\textbf{F1-score} \triangleq (2\cdot\texttt{Precision}\cdot\texttt{Recall})/(\texttt{Precision + Recall})$ is the harmonic mean of \texttt{Precision} and \texttt{Recall}.}

After account features and time-series features were produced, these two feature groups were used individually and as a combination in different models. Our overall transaction-based detection workflow is designed \hg{as follows}.

\textbf{Train-test split:} the dataset and their feature groups were split into a training set (80\%) and a test set (20\%). The former is used for training a  detection model, while the later is used to evaluate the trained model.

\textbf{Data sampling:} the Ponzi applications only occupy 6\% of total applications in our dataset. 
Therefore, we applied data sampling techniques to balance our dataset.  
We adopted the well-known oversampling method Borderline-SMOTE~\cite{han2005borderline} to generate new Ponzi instances that have more than half of the K nearest neighbours being non-Ponzi applications. That helps to enhance the existence of Ponzi applications that are more likely to be misclassified as they are located near the border of the two classes. 

\textbf{Model training:} the $K$-fold cross-validation method was used to train a classifier on the training set. In our experiment, we only set $K$ = 5, which is lower than common practice in the literature because our dataset is small.

\textbf{Model evaluation:} a trained model was used to classify the applications in the unseen test dataset. To guarantee the reliability of our experiment, we repeated the experiment process 50 times, and the final result was obtained by taking the average. It is worth mentioning that the same hyperparameters were used for the same models for a fair comparison.

\subsection{Experimental Results}
\label{subsec:experiments}

We conducted \textit{three} experiments to demonstrate the advantages of our proposed time-series features.

\textbf{Experiment 1 (Feature sets and detection models evaluation).} In this experiment, we aimed to evaluate the \hg{effectiveness} 
of our proposed feature list while applying it across diverse machine learning models.
As already mentioned, most of the previous studies used either opcode features or both opcode features and account features to build their detection models. Only a few works attempted a transaction-based approach (without using opcode features) separately \cite{chen2018detecting,jung2019data}.  
To show advantages over previous studies, we rerun their approaches on our dataset as the baselines. However, their codes and feature data have not been released to the community, so we re-implemented those models based on the descriptions in their papers, including feature lists and machine learning models. 

In this experiment, we first evaluated our feature sets corresponding to 
different time intervals ($T$ = 12, 24, 48 hours). 
\hg{We also tested with three feature sets: ACC consists of account features only, TS consists of time-series features only, and ACC-TS consists of both.} A comparison of various metrics between our feature sets and the baselines including \hg{the} feature set from Chen \emph{et al.}~\cite{chen2018detecting} (Appendix~\ref{Ap:chenfeature}) and from Jung \emph{et al.}~\cite{jung2019data}  (Appendix~\ref{Ap:jungfeature}) is provided in Table~\ref{tab:experiment_results}. 
Finally, we evaluated \hg{the} detection performance of different machine learning models using our ACC-TS feature sets \hg{(see Table~\ref{tab:model_comparision})}. 

\vspace{-10pt}
\begin{table}[htb!]
\centering
\caption{\hg{Effectiveness of the new feature set}. 
\hg{The asterisk `*' indicates that our feature list outperforms both baselines~\cite{chen2018detecting,jung2019data}}.}
\label{tab:experiment_results}
\begin{tabular}{|c|c|c|c|c|c|c|}
\hline
\textbf{Features Set}                                                    & \textbf{\hg{Number of Features}}         & \textbf{Model} & \textbf{Accuracy} & \textbf{Precision} & \textbf{Recall} & \textbf{F1}    \\ \hline
Jung \emph{et al.}~\cite{jung2019data}                                                          & 18                   & RF             & 0.966             & \textbf{0.837}     & 0.604           & 0.694          \\ \hline
Chen \emph{et al.}~\cite{chen2018detecting}                                                          & 7                    & XGB            & 0.942             & 0.587              & 0.456           & 0.499          \\ \hline
\multirow{2}{*}{ACC}                                                     & \multirow{2}{*}{29}  & RF             & 0.961             & 0.670              & 0.823$^*$            & 0.733$^*$          \\ \cline{3-7} 
&                      & XGB            & 0.965             & 0.700              & 0.835$^*$           & 0.756$^*$          \\ \hline
\multirow{2}{*}{TS}                                                      & \multirow{2}{*}{516} & RF             & 0.957             & 0.638              & 0.813$^*$           & 0.710$^*$          \\ \cline{3-7} 
                                                                         &                      & XGB            & 0.962             & 0.681              & 0.830$^*$           & 0.743$^*$          \\ \hline
\multirow{2}{*}{\begin{tabular}[c]{@{}c@{}}ACC-TS\\ 12 Hrs\end{tabular}} & \multirow{2}{*}{545} & RF             & 0.965             & 0.706              & 0.826$^*$           & 0.755$^*$         \\ \cline{3-7} 
                                                                         &                      & XGB            & 0.972$^*$             & 0.752              & 0.856$^*$           & 0.797$^*$          \\ \hline
\multirow{2}{*}{\begin{tabular}[c]{@{}c@{}}ACC-TS\\ 24 Hrs\end{tabular}} & \multirow{2}{*}{545} & RF             & 0.967$^*$             & 0.691              & 0.840$^*$           & 0.751$^*$          \\ \cline{3-7} 
                                                                         &                      & XGB            & \textbf{0.974}$^*$    & 0.743             & \textbf{0.887}$^*$  & \textbf{0.802}$^*$ \\ \hline
\multirow{2}{*}{\begin{tabular}[c]{@{}c@{}}ACC-TS\\ 48 Hrs\end{tabular}} & \multirow{2}{*}{545} & RF             & 0.969$^*$             & 0.704              & 0.816$^*$           & 0.748$^*$          \\ \cline{3-7} 
                                                                         &                      & XGB            & 0.973$^*$             & 0.733              & 0.854$^*$           & 0.782$^*$          \\ \hline
\end{tabular}
\end{table}

\hg{We} note that the \texttt{F1-score} we obtained for 
\hg{XGBoost} 
is 
\hg{close} to what was reported in~\cite{chen2018detecting} (49.9\% versus 44\%). However, we were unable to reproduce the very high scores reported in~\cite{jung2019data} for Random Forest. 
The authors of~\cite{chen2021sadponzi} also encountered the same issue: they re-implemented the approach in~\cite{jung2019data} and achieved similar \texttt{F1-score} (69.4\%) as ours (68.8\%). This could be due to the fact that in both~\cite{chen2021sadponzi} and our paper, we started from the same dataset of 1395 Ponzi and non-Ponzi schemes, while in~\cite{jung2019data}, the authors used a different dataset that includes 3203 non-Ponzi addresses. Unfortunately, 
\hg{their paper doesn't provide enough detail} on how these addresses were collected and hence, we were not able to recreate their dataset. We also noticed that although the bytecode size (\texttt{size\_info}) was created from the smart contract \textit{bytecode}, it was listed among the top eight important \textit{transaction}-based features listed in~\cite[Table~2]{jung2019data}. This makes their detection model depend on the contract code as well and therefore is susceptible to opcode obfuscation techniques~\cite{chen2021sadponzi,BiAn}. In our experiment, to reproduce their transaction-based model, we ignored this irrelevant (contract-code-based) feature \texttt{size\_info}. 

    


\hg{As shown in Table~\ref{tab:experiment_results}}, the detection models using our ACC-TS 24 Hrs feature set improved the \texttt{F1-score} of the models by Jung \emph{et al.}~\cite{jung2019data} and Chen~\emph{et al.}~\cite{chen2018detecting} by 5.7\% and 30.3\%, respectively. 
It is also clear that using both account and time-series feature 
\hg{lead to} better \texttt{Accuracy}, \texttt{Precision}, \texttt{Recall}, and \texttt{F1-score} \hg{compared to 
using individual type of features.} Note that when using time-series features (TS) alone, these models already yielded higher \texttt{F1-scores} than the baselines. 
From Table~\ref{tab:model_comparision}, we observed that tree-based classifiers were more efficient in Ponzi detection than other algorithms. More specifically, 
RF, XGB, and LGBM achieved better \texttt{Accuracy}, \texttt{Precision}, and \texttt{F1-score} values than other classifiers across different ACC-TS feature sets. Among tree-based models, LGBM with the ACC-TS 12 Hrs features achieved the best \texttt{F1-score}, \texttt{Accuracy}, and \texttt{Precision}. Last but not least, our models achieved 11\% higher F1-score for RF compared to~\cite{jung2019data} and 30\% higher F1-score for XGB compared to~\cite{chen2018detecting}.

\vspace{-15pt}
\begin{table}[]
\centering
\caption{\hg{Performance comparison among different models and feature sets.} 
The LGBM model with ACC-TS 12 Hrs features achieved the highest \texttt{F1-score}.}
\label{tab:model_comparision}
\begin{tabular}{|c|c|c|c|c|c|c|}
\hline
\textbf{Features Set}& \textbf{Number of Features}& \textbf{Model} & \textbf{Accuracy} & \textbf{Precision} & \textbf{Recall} & \textbf{F1}    \\ \hline
\multirow{5}{*}{ACC} & \multirow{5}{*}{29} & SVM  & 0.894 & 0.378 & 0.829 & 0.510 \\\cline{3-7} 
                                  &                     & KNN  & 0.898 & 0.388 & 0.875 & 0.532 \\\cline{3-7} 
                                  &                     & RF   & 0.961 & 0.670 & 0.823 & 0.733 \\\cline{3-7} 
                                  &                     & XGB  & 0.965 & 0.700 & 0.835 & 0.756 \\\cline{3-7} 
                                  &                     & LGBM & 0.967 & 0.717 & 0.823 & 0.760 \\ \hline
\multirow{5}{*}{\begin{tabular}[c]{@{}c@{}}ACC-TS\\ 12 Hrs\end{tabular}} & \multirow{5}{*}{545} & SVM            & 0.828             & 0.282              & 0.964           & 0.432          \\ \cline{3-7} 
                                                                         &                      & KNN            & 0.899             & 0.392              & 0.928           & 0.547          \\ \cline{3-7} 
                                                                         &                      & RF             & 0.965             & 0.706              & 0.826           & 0.755          \\ \cline{3-7} 
                                                                         &                      & XGB            & 0.972             & 0.752              & 0.856           & 0.797          \\ \cline{3-7} 
                                                                         &                      & LGBM           & \textbf{0.975}    & \textbf{0.779}     & 0.867           & \textbf{0.817} \\ \hline
\multirow{5}{*}{\begin{tabular}[c]{@{}c@{}}ACC-TS\\ 24 Hrs\end{tabular}} & \multirow{5}{*}{545} & SVM            & 0.845             & 0.281              & \textbf{0.973}  & 0.433          \\ \cline{3-7} 
                                                                         &                      & KNN            & 0.898             & 0.368              & 0.918           & 0.521          \\ \cline{3-7} 
                                                                         &                      & RF             & 0.967             & 0.691              & 0.840           & 0.751          \\ \cline{3-7} 
                                                                         &                      & XGB            & 0.974             & 0.743              & 0.887           & 0.802          \\ \cline{3-7} 
                                                                         &                      & LGBM           & \textbf{0.975}    & 0.768              & 0.878           & 0.812          \\ \hline
\multirow{5}{*}{\begin{tabular}[c]{@{}c@{}}ACC-TS\\ 48 Hrs\end{tabular}} & \multirow{5}{*}{545} & SVM            & 0.824             & 0.242              & 0.950           & 0.380          \\ \cline{3-7} 
                                                                         &                      & KNN            & 0.886             & 0.326              & 0.922           & 0.476          \\ \cline{3-7} 
                                                                         &                      & RF             & 0.969             & 0.704              & 0.816           & 0.748          \\ \cline{3-7} 
                                                                         &                      & XGB            & 0.973             & 0.733              & 0.854           & 0.782          \\ \cline{3-7} 
                                                                         &                      & LGBM           & 0.974             & 0.737              & 0.854           & 0.784          \\ \hline
\end{tabular}
\end{table}
\vspace{-10pt}

\begin{figure}[htb!]
\centering 
\includegraphics[width=11cm]{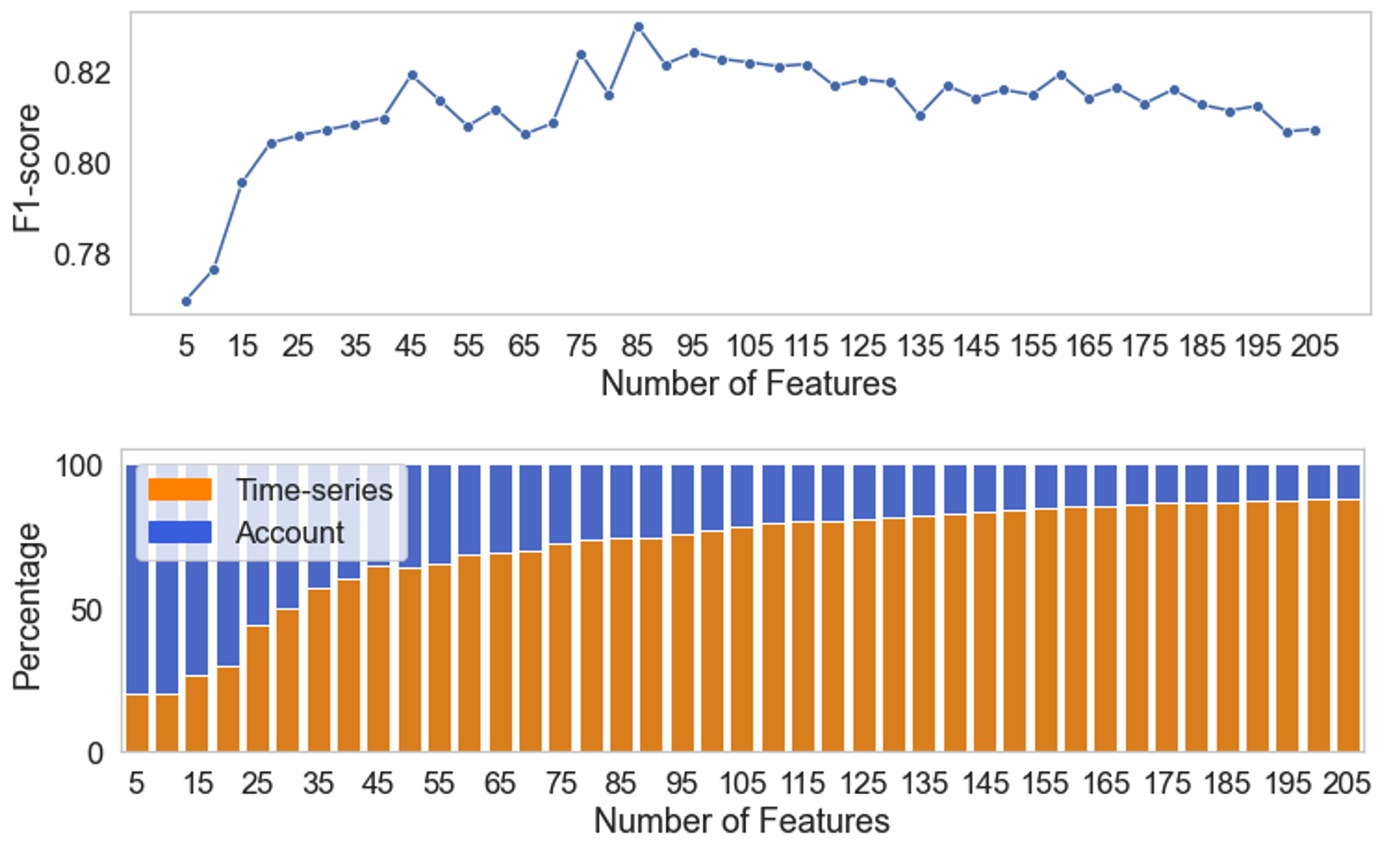}
\caption{LGBM's performance when using the 
most important features (top sub-figure) and the percentages of time-series features among these top features (bottom sub-figure). The \texttt{F1-score} value increases as more time-series features are used in the feature list, demonstrating the effectiveness of using time-series features.}
\label{fig:feature_importance}
\vspace{-10pt}
\end{figure}

\textbf{Experiment 2 (Contribution of time-series features).} Next, we investigate how much the newly proposed time-series features have contributed to LGBM's performance, \hg{which was the best performing model}. To do this, we first retrieve the list of feature importance from \hg{LGBM} 
in the previous experiment. The \textit{importance} of a feature in the LGBM model is defined to be the number of times this feature is used to split the data across all decision trees. In LGBM, an effective feature selection technique, namely Exclusive Feature Bundling (EFB), has been adopted to reduce the number of features without affecting the model's performance. We find that only 205/545 features (516 time-series features and 29 account features) had been used at least once to build a tree in the LGBM detection model. More specifically, these 205 important features consist of 176 time-series features and 29 account features. We sorted these 205 features in descending order of importance. After that, we conducted a detection with the LGBM model using only the $5, 10, 15,\ldots,205$ most important features among the 205. 
The experimental results shown in Fig.~\ref{fig:feature_importance} demonstrate how the \texttt{F1-score} values of the prediction were increased as more time-series features were added in the model.
We can also observe in the bottom sub-figure that from the top 5 onward, time-series features start to appear. For example, the top 30 contains 15 account features and 15 time-series features.

As can be seen from Fig.~\ref{fig:feature_importance}, the \texttt{F1-score} sharply increases when we increase the number of features, especially at the beginning. According to the F1-score in Fig.~\ref{fig:feature_importance}, the value exceeds 0.8 when at least 20 features out of 205 most important features are used, reaching a peak at 0.83 with the top 85 features which consist of 63 time-series features and 22 account features. Then, the \texttt{F1-score} values fluctuate around 0.81 as the number of features further increases. As shown in the bottom sub-figure of Fig.~\ref{fig:feature_importance}, the percentage of time-series features in the important feature list increases in the same direction as \texttt{F1-score}. It proves that our proposed time-series features have significantly contributed to the best-performing model (LGBM) in Experiment 1.

According to above experiment, we refined the ACC-TS 12 Hrs feature set by selecting top 85 important features of the best model (see Appendix~\ref{Ap:refinefeatures}), labelled as RF-ACC-TS 12 Hrs. To improve the detection performance, we ran all detection models with selected features instead of the full set of features. Table~\ref{tab:feature_refinement_experiment} show the detection performance results between models that use ACC-TS 12 Hrs and RF-ACC-TS 12 Hrs feature sets. Remarkably, the models using RF-ACC-TS 12 Hrs features show an average improvement of 2.1\%, 5.5\%, 1.6\%, 5.3\% for \texttt{Accuracy}, \texttt{Precision}, \texttt{Recall}, and \texttt{F1-score}, respectively.

\textbf{Experiment 3 (Detecting a new type of Ponzi).} To verify whether 
our classification model using the proposed feature list can detect a new Ponzi type, we conducted the third experiment using the LGBM model as follows. 
The key idea is to train our model on some types of Ponzi schemes and test it on another type of Ponzi schemes to see if it can still accurately detect these schemes.

\begin{table}[htb]
\centering
\caption{Our new list of 85 
features (RF-ACC-TS) completely outperformed the list of originally proposed 545 features (ACC-TS) for all metrics. The `time' column 
measures the training and prediction time (in seconds) when using different sets of features.}
\label{tab:feature_refinement_experiment}
\begin{tabular}{|c|c|c|c|c|c|c|c|}
\hline
\textbf{Features Set}                                                       & \textbf{NoF}         & \textbf{Model} & \textbf{Accuracy} & \textbf{Precision} & \textbf{Recall} & \textbf{F1}    & \textbf{\begin{tabular}[c]{@{}c@{}}Time\\ (Seconds)\end{tabular}} \\ \hline
\multirow{5}{*}{\begin{tabular}[c]{@{}c@{}}ACC-TS\\ 12 Hrs\end{tabular}}    & \multirow{5}{*}{545} & SVM            & 0.828             & 0.282              & 0.964           & 0.432          & 0.359                   \\ \cline{3-8} 
                                                                            &                      & KNN            & 0.899             & 0.392              & 0.928           & 0.547          & 0.094                   \\ \cline{3-8} 
                                                                            &                      & RF             & 0.964             & 0.696              & 0.822           & 0.749          & 1.439                   \\ \cline{3-8} 
                                                                            &                      & XGB            & 0.972             & 0.752              & 0.856           & 0.797          & 2.437                   \\ \cline{3-8} 
                                                                            &                      & LGBM           & 0.975             & 0.779              & 0.867           & 0.817          & 6.493                   \\ \hline
\multirow{5}{*}{\begin{tabular}[c]{@{}c@{}}RF-ACC-TS\\ 12 Hrs\end{tabular}} & \multirow{5}{*}{85}  & SVM            & \textbf{0.910}    & \textbf{0.425}     & \textbf{0.967}  & \textbf{0.586} & \textbf{0.111}          \\ \cline{3-8} 
                                                                            &                      & KNN            & \textbf{0.909}    & \textbf{0.422}     & \textbf{0.963}  & \textbf{0.582} & \textbf{0.063}          \\ \cline{3-8} 
                                                                            &                      & RF             & \textbf{0.973}    & \textbf{0.767}     & \textbf{0.855}  & \textbf{0.804} & \textbf{0.849}          \\ \cline{3-8} 
                                                                            &                      & XGB            & \textbf{0.973}    & \textbf{0.770}     & \textbf{0.857}  & \textbf{0.803} & \textbf{0.715}          \\ \cline{3-8} 
                                                                            &                      & LGBM           & \textbf{0.977}    & \textbf{0.795}     & \textbf{0.876}  & \textbf{0.830} & \textbf{3.076}          \\ \hline
\end{tabular}
\end{table}

We first removed all applications for each known type of Ponzi schemes mentioned in Section~\ref{subsec:ponzi} from our training set. The removed applications were then used in a test set for testing the trained model's new Ponzi detection ability. However, we only removed each of the three Ponzi types (waterfall schemes, tree-shaped schemes, and handover schemes) or all three and not the chain-shaped schemes, which account for 86\% of all the Ponzi schemes in the dataset. If we remove all chain-shaped schemes, 
the number of Ponzi samples becomes too small to learn the scam's behaviours.
Furthermore, various test sets with different scam rates were used to test our model in different situations, e.g., with a full-scam test set (100\% scams), a balance test set (50\% scams), and a few-scam test set (6\% scams, similar to our entire dataset's scam rate). Due to the lack of Ponzi (P) applications, we can only decrease the scam rate by increasing the number of non-Ponzi (non-P) applications in test sets. 

\begin{table*}[htb!]
\centering
\caption{The outcomes of Experiment 3 (Detecting a new type of Ponzi). All applications of each Ponzi type 
were removed from our training set. These applications were then used only for testing. 
We also experimented with test sets of different Ponzi rates (100\%, 50\%, and 6\%).
New Ponzi types were successfully detected with high accuracy, demonstrating the capability of our proposed feature list and detection models.}
\label{tab:detection_ability_experiment_results}
\begin{tabular}{|c|c|c|c|c|c|c|c|}
\hline
\multicolumn{1}{|c|}{\textbf{Test scheme}} & \multicolumn{1}{c|}{\textbf{Scam rate}} & \textbf{\#P} & \textbf{\#non-P} & \multicolumn{1}{c|}{\textbf{Accuracy}} & \multicolumn{1}{c|}{\textbf{Precision}} & \multicolumn{1}{c|}{\textbf{Recall}} & \multicolumn{1}{c|}{\textbf{F1-score}} \\ \hline
\multirow{3}{*}{Waterfall}                    & 100\%                                   & 4                & 0                    & 0.91                                   & 1.0                                     & 0.91                                 & 0.94                                   \\ \cline{2-8} 
                                              & 50\%                                    & 4                & 4                    & 0.94                                   & 0.98                                    & 0.89                                 & 0.93                                   \\ \cline{2-8} 
                                              & 6\%                                     & 4                & 62                   & 0.97                                   & 0.79                                    & 0.89                                 & 0.83                                   \\ \hline
\multirow{3}{*}{Tree-shaped}                  & 100\%                                   & 1                & 0                    & 1.0                                    & 1.0                                     & 1.0                                  & 1.0                                    \\ \cline{2-8} 
                                              & 50\%                                    & 1                & 1                    & 0.99                                   & 0.99                                    & 1.0                                  & 0.99                                   \\ \cline{2-8} 
                                              & 6\%                                     & 1                & 15                   & 0.98                                   & 0.87                                    & 1.0                                  & 0.91                                   \\ \hline
\multirow{3}{*}{Handover}                     & 100\%                                   & 1                & 0                    & 0.97                                   & 0.97                                    & 0.97                                 & 0.97                                   \\ \cline{2-8} 
                                              & 50\%                                    & 1                & 1                    & 0.97                                   & 0.94                                    & 0.95                                 & 0.94                                   \\ \cline{2-8} 
                                              & 6\%                                     & 1                & 15                   & 0.98                                   & 0.80                                    & 0.94                                 & 0.85                                   \\ \hline
\multirow{3}{*}{All of above}                 & 100\%                                   & 6                & 0                    & 0.92                                   & 1.0                                     & 0.92                                 & 0.95                                   \\ \cline{2-8} 
                                              & 50\%                                    & 6                & 6                    & 0.94                                   & 0.98                                    & 0.91                                 & 0.93                                   \\ \cline{2-8} 
                                              & 6\%                                     & 6                & 94                   & 0.98                                   & 0.80                                    & 0.91                                 & 0.84                                   \\ \hline
\end{tabular}
\end{table*}

The results, demonstrated in Table~\ref{tab:detection_ability_experiment_results}, indicate that the detection model can detect over 89\% of actual new Ponzi applications in a given test set (greater than 89\% of \texttt{Recall} value in most cases). Moreover, the \texttt{Precision} value is approximately 80\% even in 
test sets \hg{with very few scams}, and the model also achieved \texttt{F1-score} at least 90\% in all cases. Analysing the top 30 important features exported from the trained model, we notice that more than 80\% of features in the list are aggregated from \textit{transaction volume}, \textit{investment and payment activities}, and \textit{application balance}  with 50\% of them being time-series features. It is not surprising since those features help discriminate between Ponzi and non-Ponzi applications, as clearly shown in Section~\ref{sec:data_analysis}. That confirms the ability of the transaction-based approach to detect a new Ponzi and the importance of the time factor.
Although the dataset we use (from \cite{chen2021sadponzi}) is not ideal in the sense that there are very few Ponzi applications of types other than the chain-shaped, which may affect the reliablity of our third experiment, the outcome still gives strong evidence that a completely 
new Ponzi type can be detected.

\section{Conclusions}
\label{subsec:conclusion}
\vspace{-5pt}

In this study, we proposed a robust method for detecting Ponzi schemes in Ethereum using only the transaction data, which is harder \hg{and more costly} to be manipulated by scammers. We proposed a list of effective features that reflect the scam natures,  extracted from a careful analysis of the Ponzi and non-Ponzi schemes, in order to improve the detection accuracy of the transaction-based approach. More specifically, our analysis showed that some characteristics of a Ponzi application depend on time and should be captured by time series, representing the application's behaviours and activities throughout its lifetime. As such, we introduced a list of novel time-series features which help to significantly improve various performance metrics compared to the existing transaction-based approaches.

\hg{There are several open problems left for future research.}
\textit{First}, although we have considerably increased the detection accuracy of transaction-based detection tools, there is still room for future improvement. Specifically, one open problem is to find more effective statistical measures to capture the global information of time series. \textit{Second}, it is desirable to collect more data to extend the ground-truth dataset for Ponzi applications, which will help to train the detection models more effectively. Moreover, we can test our approaches on other popular algorithms that work effectively on big data such as Artificial Neural Networks~\cite{wang2003artificial} or Recurrent Neural Networks~\cite{medsker2001recurrent}. \textit{Finally}, blockchain scams are becoming more sophisticated. Scammers may use multiple smart contracts or smart contracts from a third party as an additional service, making the picture much more complicated. In such cases, the application's transactions might not be enough to perform fraud detection. Detecting such sophisticated scams remains a big challenge \hg{for future research.}

%
%
%
%
{\footnotesize \bibliographystyle{splncs04}
\bibliography{main}}
\appendix

\section{Ponzi Types}
\label{app:Ponzi_types}
\vspace{-10pt}

Bartoletti~\emph{et al.}~\cite{BARTOLETTI2020259} defined the following types of Ponzi schemes on blockchain.

\textbf{Chain-shaped schemes} use a linear money distribution mechanism. These schemes often commit to paying investors a multiple, e.g., double, of their original investments. Each new investor joining the scheme is appended to a payment list in their order of arrival. 
Each investor in the list is paid in full with their promised amount whenever the accumulated fund (minus some commission fee) is sufficient. 
These schemes will collapse when the investment becomes too large to fulfill and the waiting time of late comers grows. 

\textbf{Tree-shaped schemes} use a tree structure to manage the money redistribution, in which an inviter is a parent node, and the invitees are their children. Once a new investor joins the scheme, his investment is split and distributed among the ancestors: the nearer an ancestor is, the more he will receive. In this type of Ponzi scheme, investors cannot guess how much they will gain because their profit depends on how many users they and their descendants can invite and also how much these users pay. Similar to other schemes, tree-shaped Ponzi collapses when there are no or too few users joining.

\textbf{Handover schemes}, like Chain-shaped schemes, also use a linear payment list. However, instead of gathering newcomers' investments, these schemes require the entry toll, which increases every time a new user joins this scheme. At a time, only one user is invited by the last user in the list, and the new entry toll is paid entirely to the inviter to make an instant profit. Once the inviter is paid, he hands the privilege over to the following user who just came. 

\textbf{Waterfall schemes} are similar to chain-shaped schemes in payment order but different in money distribution logic. Every new investment is distributed along the list of existing investors from the first to the last or until the fund is exhausted. This first-join-first-receive logic implies that the later joining investors are less likely to reap any profit. 

\section{Account feature list}
\label{Ap:afeatures}
The list of 29 different \textit{account features}, which are used to represent the general characteristics of an application, are given below.

\subsection{List of features from Chen \emph{et al.}~\cite{chen2018detecting}}
\label{Ap:chenfeature}
\begin{itemize}
    \item \textbf{know\_rate}: the proportion of participants who have invested before payment.  
    \item \textbf{balance}: the balance of a smart contract.
    \item \textbf{num\_in\_txs}: total number of transactions sent to a contract (investments).         
    \item \textbf{num\_out\_txs}: total number of transactions sent out from a contract (payments).
    \item \textbf{difference\_idx}: the difference measurement of counts between payment and investment. 
    \item \textbf{paid\_rate}: the proportion of investors who received at least one payment.    
    \item \textbf{max\_pay}: maximum number of payment to a participant. 
\end{itemize}

\subsection{List of features from Jung \emph{et al.}~\cite{jung2019data}.}
\label{Ap:jungfeature}
\begin{itemize}
    \item \textbf{num\_in\_txs} (abv): total number of transactions sent to a contract (investments).         
    \item \textbf{num\_out\_txs} (abv): total number of transactions sent out from a contract (payments).
    \item \textbf{total\_inv\_amt}: total ETH amount sent to a contract (total investment amount).                                    
    \item \textbf{total\_pay\_amt}: total ETH amount sent out from a contract (total payment amount).  
    \item \textbf{avg\_inv\_amt}: the average of ETH amounts sent to a contract.
    \item \textbf{avg\_pay\_amt}: the average of ETH amounts sent out from a contract.
    \item \textbf{dev\_inv\_amt}: the standard deviation of ETH amounts sent to a contract.
    \item \textbf{dev\_pay\_amt}:  the standard deviation of ETH amounts sent out from a contract.
    \item \textbf{avg\_time\_btw\_txs}: the average of time distance between two consecutive transactions.
    \item \textbf{life\_time}: a contract lifetime.
    \item \textbf{gini\_amt\_in}: the Gini coefficient of total ETH amount sent to a contract.   
    \item \textbf{gini\_amt\_out}: the Gini coefficient of total ETH amount received from a contract.                
    \item \textbf{avg\_time\_btw\_txs}: the average of time distance between two consecutive transactions.
    \item \textbf{overlap\_addr}: the number of addresses that invested and be paid by contract.
    \item \textbf{gini\_time\_in}: the Gini coefficient of number of transactions sent to a contract.
    \item \textbf{gini\_time\_out}: the Gini coefficient of number of transactions sent out from a contract.
    \item \textbf{num\_inv\_acc}: number of distinct account addresses that send ETH to a contract.
    \item \textbf{num\_pay\_acc}: number of distinct account addresses that send ETH to a contract.
\end{itemize}
* We ignored the irrelevant (contract-code-based) feature \texttt{size\_info}.
\subsection{Other Account Features}
\begin{itemize}
    \item \textbf{balance\_rate}: ratio between balance and total investment.
    \item \textbf{payment\_time}: ratio between balance and total investment.
    \item \textbf{num\_all\_txs}: total number of tranasctions.
    \item \textbf{num\_in\_txs}: total number of transactions sent to a contract.
    \item \textbf{num\_out\_txs}: total number of transactions sent out from a contract.
    \item \textbf{pay\_skewness}: a payment skewness.                                                    
\end{itemize}

\section{Time-series list}
\label{Ap:ifeatures}
Below is the list of 43 different time series we used to represent the change of information associated with an application throughout its lifetime in different aspects. Those time series were derived from basic transaction information. We then grouped them by the information from which they were created. All of these data depend on time, e.g., which day they were measured.
\subsection{ETH value:}
\begin{itemize}
    \item \textbf{balance}: the amount of ETH in a contract.
    \item \textbf{profit\_and\_loss}: subtraction of total investments (profit) and total payments (loss) of a contract.
    \item \textbf{loss}: total ETH amounts that a contract pays to its participants.
    \item \textbf{loss\_by\_contract}: total ETH amount sent from a contract to other contracts.
    \item \textbf{loss\_by\_person}: total ETH amount sent from a contract to the other user accounts.
    \item \textbf{loss\_from\_internal\_txs}: total ETH amount recorded by internal transactions that a contract pays to its participants.
    \item \textbf{loss\_from\_normal\_txs}: total ETH amount recorded by external transactions that a contract pays to its participants.
    \item \textbf{profit}: total ETH amount that the contract received from its participants.
    \item \textbf{profit\_by\_contract}: total ETH amount that the contract received from other contracts.
    \item \textbf{profit\_by\_person}: total ETH amount that the contract received from other user accounts.
    \item \textbf{profit\_from\_internal\_txs}: total ETH amount recorded by internal transactions that a contract pays received from its participants.
    \item \textbf{profit\_from\_normal\_txs}: total ETH amount recorded by external transactions that a contract pays received from its participants.
\end{itemize}

\subsection{Transaction:}
\begin{itemize}
    \item \textbf{total\_txs}: total number of transactions.
    \item \textbf{total\_internal\_txs}: total number of internal transactions
    \item \textbf{total\_in\_coming\_txs}: total number of transactions sent to a contract.
    \item \textbf{total\_in\_coming\_internal\_txs}: total number of internal transactions sent to the contract.
    \item \textbf{total\_in\_coming\_normal\_txs}: total number of external transactions sent to the contract.
    \item \textbf{total\_normal\_txs}: total number of external transactions
    \item \textbf{total\_out\_going\_txs}: total number of transactions sent from a contract.
    \item \textbf{total\_out\_going\_internal\_txs}: total number of internal transactions sent from a contract.
    \item \textbf{total\_out\_going\_normal\_txs}: total number of external transactions sent from a contract.
\end{itemize}

\subsection{Participant address:}
\begin{itemize}
    \item \textbf{total\_unique\_addresses}: total number of distinct participants (addresses) of a contract.
    \item \textbf{total\_unique\_in\_coming\_addresses}: total number of distinct participants who sent transactions to a contract.
    \item \textbf{total\_unique\_in\_coming\_addresses\_from\_internal}: total number of distinct participants who sent internal transactions to a contract.
    \item \textbf{total\_unique\_in\_coming\_addresses\_from\_normal}:  total number of distinct participants who sent external transactions to a contract.
    \item \textbf{total\_unique\_out\_going\_addresses}: total number of distinct participants who receive transactions from a contract.
    \item \textbf{total\_unique\_out\_going\_addresses\_from\_internal}: total number of distinct participants who receive internal transactions from a contract.
    \item \textbf{total\_unique\_out\_going\_addresses\_from\_normal}: total number of distinct participants who receive external transactions from a contract.
\end{itemize}

\subsection{Calling Function\protect\footnote{ Calling functions can be retrieved from a transaction's input data}:}
\begin{itemize}
    \item \textbf{total\_unique\_calling\_function}: total number of distinct functions were called by a contract or its participants.
    \item \textbf{total\_unique\_in\_coming\_calling\_function}: total number of distinct functions called by participants.
    \item \textbf{total\_unique\_in\_coming\_calling\_function\_from\_internal}: total number of distinct functions called via internal transactions by participants.
    \item \textbf{total\_unique\_in\_coming\_calling\_function\_from\_normal}: total number of distinct functions called via external transactions by participants.
    \item \textbf{total\_unique\_out\_going\_calling\_function}: total number of distinct functions called by contracts.
    \item \textbf{total\_unique\_out\_going\_calling\_function\_from\_internal}:  total number of distinct functions called via internal transactions by contracts
    \item \textbf{total\_unique\_out\_going\_calling\_function\_from\_normal}: total number of distinct functions called via external transactions by contracts
\end{itemize}

\subsection{Participant Account Type:}
\begin{itemize}
    \item \textbf{num\_in\_coming\_txs\_from\_contract}: number of transactions sent to a contract from other contracts
    \item \textbf{num\_in\_coming\_txs\_from\_person}: number of transactions sent to a contract from other user accounts.
    \item \textbf{num\_out\_going\_txs\_to\_contract}: number of transactions sent from a contract to other contracts
    \item \textbf{num\_out\_going\_txs\_to\_person}: number of transactions sent from a contract to other user accounts.
    \item \textbf{num\_unique\_in\_coming\_contract\_address}: number of distinct contracts that sent transactions to a contract
    \item \textbf{num\_unique\_in\_coming\_person\_address}: number of distinct user accounts that sent transactions to a contract
    \item \textbf{num\_unique\_out\_going\_contract\_address}: number of distinct contracts that received transactions from a contract
    \item \textbf{num\_unique\_out\_going\_person\_address}: number of distinct user accounts that received transactions from a contract.
\end{itemize}

\section{Time-series Statistical Measures}
\label{Ap:tsmeasures}
Below are the 12 statistical measures that were used to capture the characteristics of a time series.
\begin{itemize}

\item \textbf{Mean}: Mean value of intervals  
\item \textbf{Var}: Variance value of intervals                                                          \item \textbf{ACF1}: First order of auto-correlation of the series                                    
\item \textbf{Linearity}:Strength of linearity calculated based on the coefficients of an orthogonal quadratic regression        
\item \textbf{Curvature}: Strength of curvature  calculated based on the coefficients of an orthogonal quadratic regression       
\item \textbf{Trend}: Strength of trend of a time-series based on an STL decomposition 
\item \textbf{Season}: Strength of seasonality of a time-series based on an STL
\item \textbf{Entropy}: Spectral entropy measures the “forecastability” of a time-series, where low values indicate a high signal-to-noise ratio, and large values occur when a series is difficult to forecast \item \textbf{Lumpiness}: Changing variance in remainder computed on non-overlapping windows  
\item \textbf{Spikiness}: Strength of spikiness which is variance of the leave-one-out variances of the remainder component 
\item \textbf{Fspots}: Flat spot using discretization computed by dividing the sample space of a time-series into ten equal-sized intervals, and computing the maximum run length within any single interval. 
\item \textbf{Cpoints}: The number of times a time-series crosses the mean line
\end{itemize}

\section{Refined Feature List of 85 Features}
\label{Ap:refinefeatures}
\begin{table}[]
\resizebox{\textwidth}{!}{%
\begin{tabular}{|l|l|l|l|l|l|}
\hline
\textbf{Rank} & \textbf{Feature}                            & \textbf{Type} & \textbf{Rank} & \textbf{Feature}                              & \textbf{Type} \\ \hline
1             & avg\_inv\_amt                               & ACC           & 44            & profit\_and\_loss\_entropy                    & TD            \\ \hline
2             & num\_all\_txs                               & ACC           & 45            & profit\_and\_loss\_variance                   & TD            \\ \hline
3             & balance                                     & ACC           & 46            & profit\_and\_loss\_curvature                  & TD            \\ \hline
4             & avg\_pay\_amt                               & ACC           & 47            & difference\_idx                               & ACC           \\ \hline
5             & total\_txs\_lumpiness                       & TD            & 48            & total\_in\_coming\_normal\_txs\_fspots        & TD            \\ \hline
6             & avg\_time\_btw\_txs                         & ACC           & 49            & total\_in\_coming\_internal\_txs\_acf1        & TD            \\ \hline
7             & total\_in\_coming\_txs\_lumpiness           & TD            & 50            & gini\_amt\_in                                 & ACC           \\ \hline
8             & balance\_rate                               & ACC           & 51            & total\_txs\_spikiness                         & TD            \\ \hline
9             & dev\_inv\_amt                               & ACC           & 52            & loss\_lumpiness                               & TD            \\ \hline
10            & num\_in\_txs                                & ACC           & 53            & total\_txs\_acf1                              & TD            \\ \hline
11            & paid\_rate                                  & ACC           & 54            & total\_in\_coming\_normal\_txs\_entropy       & TD            \\ \hline
12            & gini\_amt\_out                              & ACC           & 55            & gini\_time\_out                               & ACC           \\ \hline
13            & profit\_and\_loss\_acf1                     & TD            & 56            & total\_in\_coming\_normal\_txs\_mean          & TD            \\ \hline
14            & total\_internal\_txs\_lumpiness             & TD            & 57            & total\_internal\_txs\_mean                    & TD            \\ \hline
15            & dev\_pay\_amt                               & ACC           & 58            & profit\_from\_normal\_txs\_mean               & TD            \\ \hline
16            & total\_inv\_amt                             & ACC           & 59            & total\_in\_coming\_internal\_txs\_fspots      & TD            \\ \hline
17            & gini\_time\_in                              & ACC           & 60            & profit\_and\_loss\_trend                      & TD            \\ \hline
18            & payment\_time                               & ACC           & 61            & nbr\_tx\_in                                   & ACC           \\ \hline
19            & total\_in\_coming\_internal\_txs\_linearity & TD            & 62            & total\_in\_coming\_normal\_txs\_linearity     & TD            \\ \hline
20            & total\_in\_coming\_internal\_txs\_lumpiness & TD            & 63            & total\_in\_coming\_internal\_txs\_entropy     & TD            \\ \hline
21            & total\_in\_coming\_normal\_txs\_lumpiness   & TD            & 64            & total\_pay\_amt                               & ACC           \\ \hline
22            & profit\_mean                                & TD            & 65            & profit\_spikiness                             & TD            \\ \hline
23            & profit\_and\_loss\_mean                     & TD            & 66            & total\_in\_coming\_normal\_txs\_cpoints       & TD            \\ \hline
24            & total\_internal\_txs\_cpoints               & TD            & 67            & num\_inv\_acc                                 & ACC           \\ \hline
25            & total\_txs\_fspots                          & TD            & 68            & total\_internal\_txs\_spikiness               & TD            \\ \hline
26            & profit\_and\_loss\_linearity                & TD            & 69            & total\_txs\_variance                          & TD            \\ \hline
27            & total\_txs\_linearity                       & TD            & 70            & total\_in\_coming\_txs\_acf1                  & TD            \\ \hline
28            & know\_rate                                  & ACC           & 71            & num\_out\_going\_txs\_to\_contract\_lumpiness & TD            \\ \hline
29            & total\_txs\_cpoints                         & TD            & 72            & total\_in\_coming\_txs\_entropy               & TD            \\ \hline
30            & balance\_acf1                               & TD            & 73            & total\_txs\_trend                             & TD            \\ \hline
31            & total\_txs\_mean                            & TD            & 74            & profit\_linearity                             & TD            \\ \hline
32            & total\_in\_coming\_txs\_linearity           & TD            & 75            & total\_in\_coming\_normal\_txs\_spikiness     & TD            \\ \hline
33            & profit\_lumpiness                           & TD            & 76            & total\_in\_coming\_normal\_txs\_acf1          & TD            \\ \hline
34            & profit\_entropy                             & TD            & 77            & loss\_entropy                                 & TD            \\ \hline
35            & profit\_and\_loss\_fspots                   & TD            & 78            & num\_out\_going\_txs\_to\_person\_entropy     & TD            \\ \hline
36            & total\_internal\_txs\_linearity             & TD            & 79            & profit\_from\_normal\_txs\_entropy            & TD            \\ \hline
37            & profit\_acf1                                & TD            & 80            & total\_internal\_txs\_acf1                    & TD            \\ \hline
38            & total\_txs\_entropy                         & TD            & 81            & max\_pay                                      & ACC           \\ \hline
39            & balance\_mean                               & TD            & 82            & loss\_acf1                                    & TD            \\ \hline
40            & overlap\_addr                               & ACC           & 83            & total\_in\_coming\_internal\_txs\_spikiness   & TD            \\ \hline
41            & total\_in\_coming\_internal\_txs\_mean      & TD            & 84            & num\_out\_going\_txs\_to\_contract\_spikiness & TD            \\ \hline
42            & total\_internal\_txs\_entropy               & TD            & 85            & num\_out\_going\_txs\_to\_contract\_mean      & TD            \\ \hline
43            & total\_out\_going\_txs\_entropy             & TD            &               &                                               &               \\ \hline
\end{tabular}}
\end{table}

\end{document}